\documentclass[aps,pra,twocolumn,showpacs,amsmath,amssymb,preprintnumbers,superscriptaddress,10pt]{revtex4-1}

\usepackage{SIunits}
\usepackage{hyphenat}
\usepackage[bookmarks=false]{hyperref}
\usepackage{color}
\usepackage[capitalise]{cleveref}
\crefname{section}{Sec.}{Secs.}
\Crefname{section}{Section}{Sections}
\usepackage{floatrow}
\usepackage{graphicx}
\usepackage[caption=false]{subfig}
\newcommand*\diff{\mathop{}\!\mathrm{d}}

\begin{document} 

\title{Quantum entropy model of an integrated QRNG chip}

\author{Ga\"etan~Gras}
\email{gaetan.gras@idquantique.com}
\affiliation{ID Quantique SA, CH-1227 Carouge, Switzerland}
\affiliation{Group of Applied Physics, University of Geneva, CH-1211 Geneva, Switzerland}

\author{Anthony Martin}
\affiliation{ID Quantique SA, CH-1227 Carouge, Switzerland}

\author{Jeong Woon Choi}
\affiliation{ID Quantique SA, CH-1227 Carouge, Switzerland}

\author{F\'elix~Bussi\`eres}
\affiliation{ID Quantique SA, CH-1227 Carouge, Switzerland}

\date{\today} 

\begin{abstract}
We present the physical model for the entropy source of a quantum random number generator chip based on the quantum fluctuations of the photon number emitted by light-emitting diodes. This model, combined with a characterization of the chip, estimates a quantum min-entropy of over 0.98 per bit without post-processing. Finally, we show with our model that the performances in terms of security are robust against fluctuations over time.
\end{abstract}

\maketitle

\section{Introduction}

Random numbers are used in a wide range of applications such as gambling, numerical simulations and cryptography. Lack of a good random number generator (RNG) can have serious consequences on the security of devices and protocols \cite{dorrendorf2009,ps3, bitcoin}. Currently, many applications rely on RNGs based on a stochastic process and lack a complete security model. In order to have a sequence usable for cryptographic applications, the source of randomness must be completely unpredictable, even if a malicious adversary has a perfect description of the system \cite{kerckhoffs1883}. Quantum RNGs (QRNGs) can overcome this problem thanks to the intrinsically probabilistic nature of quantum mechanics. 
One key challenge today is to have a fully integrated QRNG device that reach mass-market deployment. Several works have been carried out toward that goal \cite{Stefanov2000,furst2010,Sanguinetti2014,Tisa2015,Khanmohammadi2015,Abellan2016,Zhang2016,amri2016,Raffaelli2018,Raffaelli2018a,bisadi2018,leone2020,Stanco2020,Imran2020}. One of them is a QNRG implementation based solely on components that are compatible with integrated electronics, namely a light-emitting diode (LED), a CMOS image sensor (CIS) and an analog-to-digital converter (ADC) \cite{Sanguinetti2014}. More precisely, this work showed that a CIS-based mobile phone camera could be used as an entropy source providing 10-bit long strings containing 5.7 bits of quantum entropy. However, this approach still required software-based randomness extraction to generate bits with close-to-maximal entropy, and a fully integrated implementation remained to be demonstrated. 

In this paper, we present a fully integrated QRNG architecture and chip implementation based on the quantum statistics of light captured by a CIS, and we present a model showing that the quantum entropy of each bit produced is close to unity without the need of randomness extraction. This architecture is used to provide small-form factor and low power consumption chips, making them suitable for mobile devices such as smartphones.

\section{Physical model}

	\subsection{Chip architecture}
	
\begin{figure}[htbp]
	\includegraphics[width=\columnwidth]{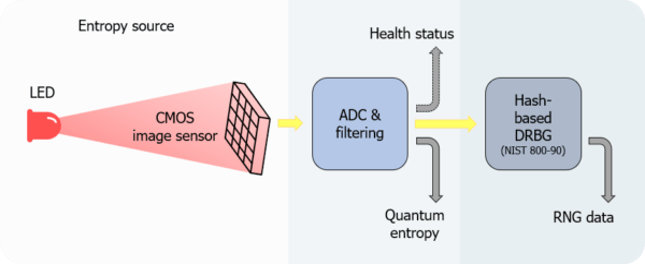}
	\caption{Schematic representation of the QRNG. All the components are embedded on a single chip.}
	\label{fig:chip_scheme}
\end{figure}

A scheme of the architecture of IDQ's QRNG chips is shown in \cref{fig:chip_scheme}. A LED is used as a continuous source of photons. As the light field emitted is highly multi-mode, the probability
distribution of the photon number is very well approximated by a Poisson distribution with mean $\mu_\text{ph}$ \cite{papen2019}. The probability to have $n$ photons emitted during a fixed time interval is given by : 
\begin{equation}
p(n,\mu_\mathrm{ph}) = \frac{\mu_\mathrm{ph}^n}{n!}\mathrm{e}^{-\mu_\mathrm{ph}}
\end{equation}

\begin{figure}
  \sidesubfloat[]{\includegraphics{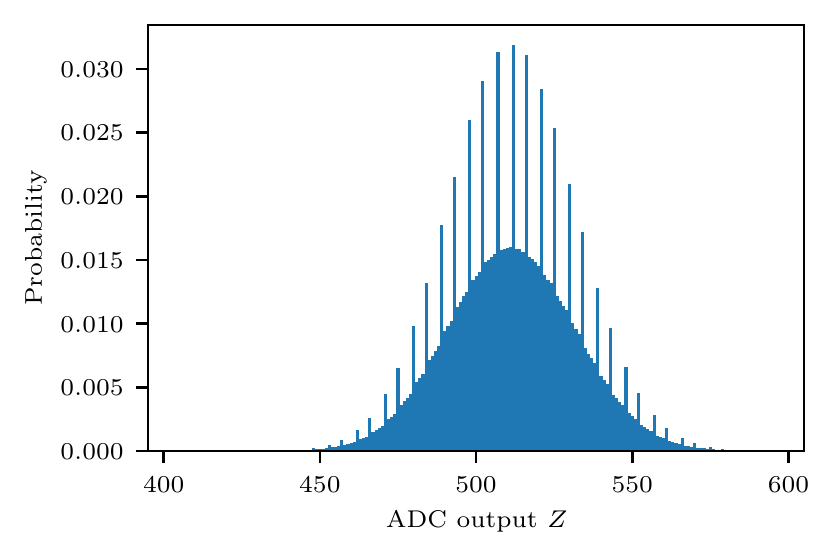}\label{sub:Z}}\\%
  \sidesubfloat[]{\includegraphics{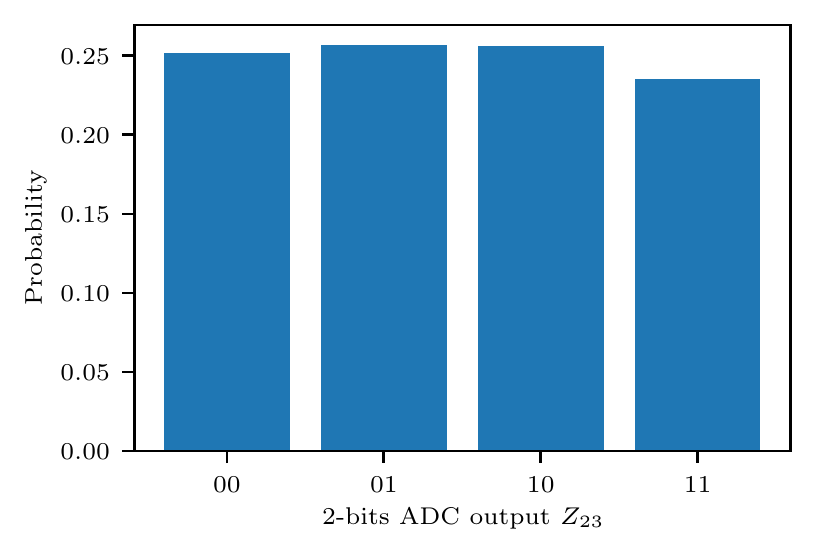}\label{sub:Z23}}%
  \caption{(a) Simulated ADC output distribution in case there is no noise, with $K=0.8192$ (obtained from the factory-given parameters of the chip). (b) Two bits probability distribution simulated from (a) giving a min-entropy per bit $H_{\min}=0.982$.}
  \label{fig:ADC_output_sim}
\end{figure}

Photons are converted into electrons by a CMOS image sensor array during the integration time of the sensor. Each pixel of the sensor has an efficiency $\eta$ (taking into account transmission losses and detection efficiencies) that may vary between them. The number of electrons $N_\mathrm{e}$ is directly correlated to the quantum fluctuations of the LED and follows a Poisson distribution with mean value $\mu_\mathrm{e} = \eta\mu_\mathrm{ph}$. We assume that pixels are independent from each other and that there is no correlation from frame to frame (these assumptions are verified in \cref{sec:tests}). After accumulation, the number of electrons is converted into a voltage, then digitized with a 10-bits ADC. We define $K$ as the gain between $N_\mathrm{e}$ and the analog-to-digital unit of the ADC. We also define two random variables $X$ and $Z$. $X$ is a continuous random variable representing the voltage value distribution at the input of the ADC and can be written  
\begin{equation}
X = KN_\text{e} + E
\end{equation}
where $E$ is the random variable associated with the classical noise (see \cref{sec:noise}). $Z$ is the random variable returned by the ADC and is defined as 
\begin{equation}
Z =\begin{cases}
0 &\text{ if } X < 0 \\
\lfloor X\rfloor &\text{ if } X \in [0;1023] \\
1023 &\text{ if } X>1023 
\end{cases}
\end{equation}
where $\lfloor . \rfloor$ is the floor operator. \Cref{sub:Z} shows a simulated distribution of $Z$ with $\mu_\mathrm{e}=625$. On this graph, we observe a normal distribution of the ADC output values, combined with a series of peaks with twice the probability. This ``pile-up" effect is due to the factor $K$ of the chip which is inferior to 1. As one electron is not enough to increase the signal by a full ADC step, two electron numbers can lead to the same ADC output making this value twice more probable, with a periodicity that goes roughly like $1/(1-K)$.


\floatsetup[figure]{style=plain,subcapbesideposition=top}

To generate entropy bits from the 10-bits ADC output $Z$, we keep the least significant bits (LSB) 2 and 3, noted $Z_{23}$. Indeed, their entropy is the more robust of all the bits against imperfections of the system. This happens because the most significant bits will be biased if $\mu_\mathrm{e}$ is not well controlled. Moreover, LSB 0 and 1 can be affected by small and uncontrolled fluctuations which are not from quantum origin, but also by the pile-up effect. By taking only LSB 2 and 3, we can easily mitigate these effects to obtain bits with a very high min-entropy $H_{\min}$ without post-processing as it can be seen in \cref{sub:Z23}. We note that this principle can be applied with ADCs of different resolution, with the right choice of bits retained to generate the entropy bits. These two bits can be used as entropy bits directly, or can be seeded to a Hash-based deterministic random bit generator (DRBG) as recommended by the National Institute of Standards and Technology (NIST) documentations (SP~800-90A) \cite{NIST_SP800_90A}.

	\subsection{Noise model}
	\label{sec:noise}
	
\begin{figure}[htbp]
	\includegraphics[width=0.9\columnwidth]{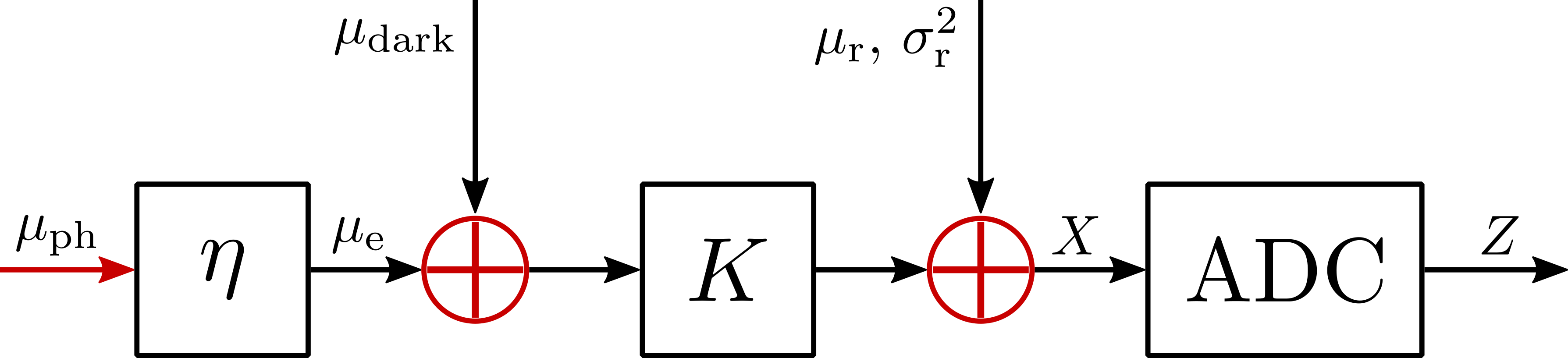}
	\caption{Schematic representation of the noise sources in the chip. Dark electrons are added to the electrons generated by the LED. The total number of electrons is converted into a voltage with a factor $K$. After conversion, noise from the readout circuit is added before the signal is digitized with the ADC.}
	\label{fig:NoiseSources}
\end{figure}

To complete our model, we need to take into account the classical noise $E$ as it can impact the security of the chip. We consider two sources of noise as shown in \cref{fig:NoiseSources}. 

First, we have a discrete source of dark electrons which are generated by another process than the absorption of a photon emitted by the LED (e.g. thermal excitation). These follow a Poisson distribution with parameter $\mu_\text{dark}$ and are added to the photo-electrons. Second, we consider a continuous source due to electronic noise in the readout circuit following a normal probability distribution $\mathcal{N}$ described by a probability density function $\Phi_{\mu_\mathrm{r},\sigma_\mathrm{r}}$ with mean $\mu_\text{r}$ and variance $\sigma^2_\text{r}$ \cite{Teranishi2012, Aguerrebere2012, Seo2015}. The probability density function $P_E$ of the classical noise is therefore a convolution of a Poisson and a normal distribution and can be written : 
\begin{equation}
P_E(e) = \sum_n p(n,\mu_\text{dark})\Phi_{\mu_\mathrm{r}+Kn,\sigma_\mathrm{r}}(e).
\label{eq:noise}
\end{equation} 
	
We assume all sources of classical noise are accessible to an adversary (called Eve). We suppose that Eve cannot change it after fabrication and characterization of the chip and that is it uncorrelated to the quantum entropy source. We then need to calculate the min-entropy of $Z_{23}$ given $E$ as it is defined in \cite{Tomamichel2011b} :
\begin{equation}
H_{\min} (Z_{23}|E) = -\log_2\left(p_{guess}\right),
\end{equation}
where
\begin{equation}
p_{guess} = \int P_E(e) \max_{z_{23}}\left(P_{Z_{23}|E=e}(z_{23})\right)\diff e
\end{equation}
is the optimal guessing probability of $Z_{23}$ given $E$. This gives the quantum min-entropy output of the chip.

\section{Experimental characterization}

In our model, we made several assumptions (photon number distribution, independence between pixels and between frame). In this section, we show results from measurements on a QRNG chip (model IDQ6MC1) to validate these assumptions.

	\subsection{Light source}
	
\begin{figure}
  \sidesubfloat[]{\includegraphics{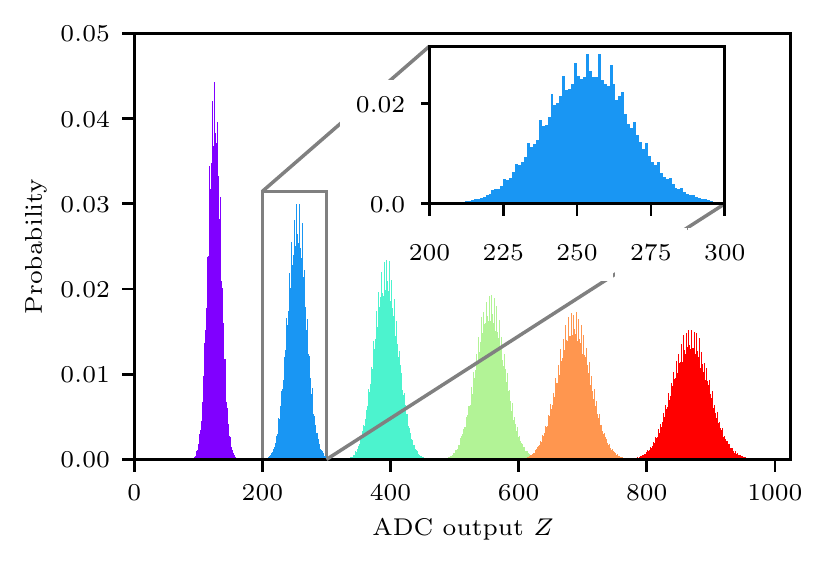}\label{sub:pow}}\\%
  \sidesubfloat[]{\includegraphics{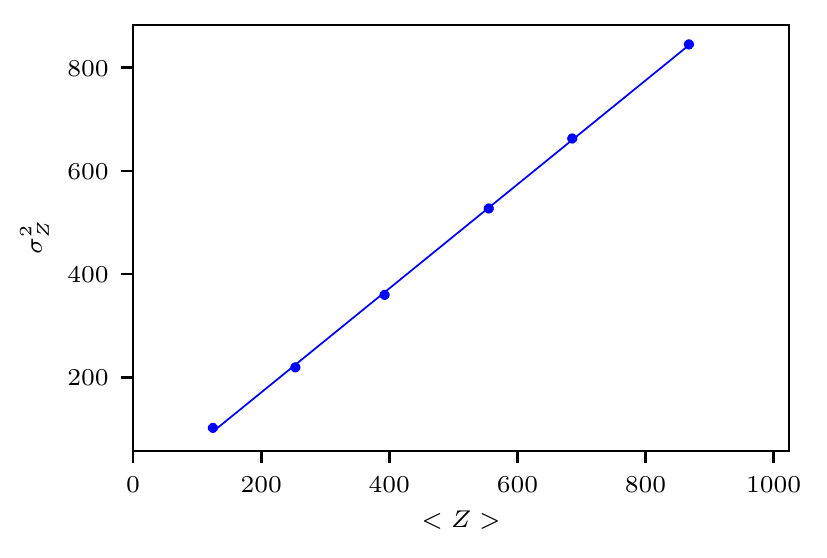}\label{sub:var}}%
  \caption{(a) ADC output distribution $Z$ given by one pixel of the array for various light intensity. (b) Variance of $Z$ versus its mean value for the distributions of (a).}
  \label{fig:VarVsMean}
\end{figure}

Firstly, we want to characterize our source in order to verify that the number of photons emitted follows a Poisson statistics. To achieve that goal, we can measure the distribution of the ADC output $Z$ for various intensities by changing the current inside the LED. Results are displayed in \cref{sub:pow}. On the plot, we can observe a pile-up effect similar to the one predicted by our model (see \cref{sub:Z}). Peaks are less prominent than in our simulations; that is due to the presence of the classical noise averaging them out. From these acquisitions, we can plot the variance of $Z$, $\sigma_Z^2$, as a function of its expected value $\left<Z\right>$ (see \cref{sub:var}). Due to the conversion factor $K$ affecting the mean value and the variance of the number of electrons differently and the offset of the ADC, we do not have $\left<Z\right>=\sigma_Z^2$ as expected from a Poisson distribution. Nevertheless, this does not affect the linear relationship between them, as we can see in \cref{sub:var}, validating the Poissonian nature of the light emitted by the LED and the transfer of this statistics to the electron number distribution.

\floatsetup[figure]{style=plain,subcapbesideposition=top}


	\subsection{Classical noise}
	
\begin{figure}[htbp]
	\includegraphics[width=\columnwidth]{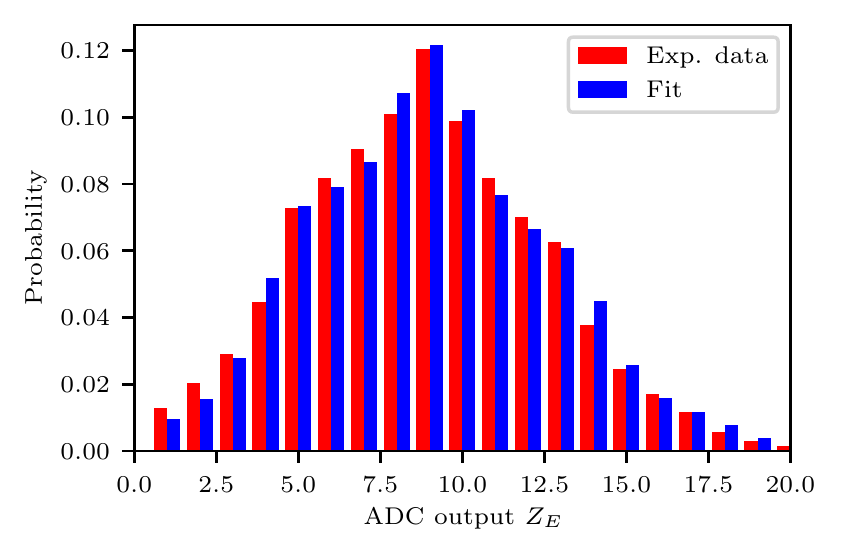}
	\caption{Noise distribution of one of the pixels.}
	\label{fig:noise_hist}
\end{figure}

We characterize the noise distribution for 4 different pixels on the array. For that, we switch off the LED and measure the distribution $Z_E$ at the output of the ADC with only classical noise. As this distribution is centered near 0 in the default settings, we adjust the ADC offset to shift it to the right by 8 ADC steps in order to see the distribution completely. The histogram of $Z_E$ is given in \cref{fig:noise_hist}. We observe a similar pile-up effect to the one observed with the LED on coming from the discrete component of $E$. We can fit this histogram with \cref{eq:noise} to extract the different parameters of the classical noise presented in \cref{tab:noise}. The value $\mu_r$ depends on the ADC offset but we can extrapolate from our measurements in order to find its value for the default settings of the chip.

\begin{table}[htbp]
\caption{Parameters of the noise distribution for 4 pixels of the CMOS image sensor. The value for $\mu_r$ was extrapolated from our measurement to find the value with the default ADC offset.}
\begin{tabular*}{0.8\columnwidth}[t]{@{\extracolsep{\fill}}cccc}
   \hline
   \hline
   Pixel label & $\mu_r$ & $\sigma_r$ & $\mu_\mathrm{dark}$ \\
   \hline
   1 & -13.6 & 0.21 & 17.2 \\
   2 & -16.8 & 0.22 & 18.0 \\
   3 & -14.4 & 0.23 & 17.2 \\
   4 & -13.6 & 0.21 & 19.0 \\
   \hline
   \hline
\end{tabular*}
\label{tab:noise}
\end{table} 

As we can see, classical noise is mainly given by dark electrons ($\mu_\mathrm{dark}>>\sigma_\mathrm{r}^2$). Moreover, the noise parameters for the 4 pixels spread across the array are quite close. We can therefore assume all the pixels will have similar noise distribution.	

	\subsection{Correlation measurements}
	\label{sec:tests}

\begin{figure}
  \sidesubfloat[]{\includegraphics{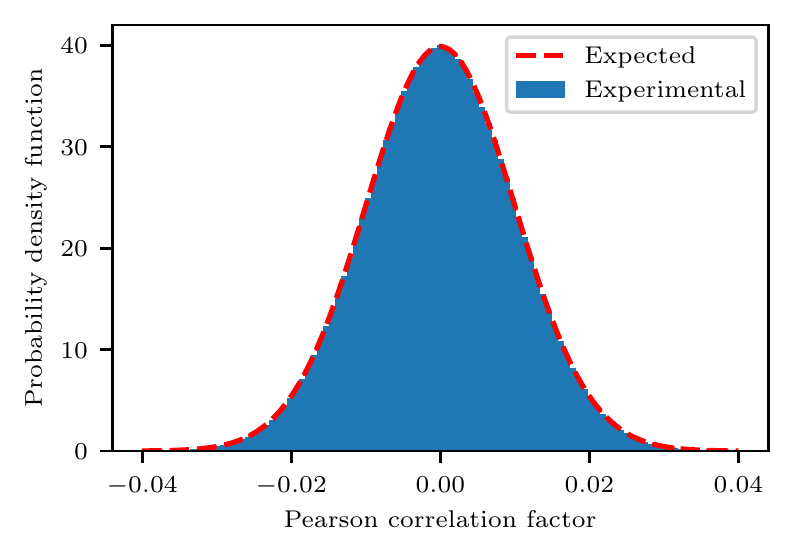}\label{sub:correlation}}\\%
  \sidesubfloat[]{\includegraphics{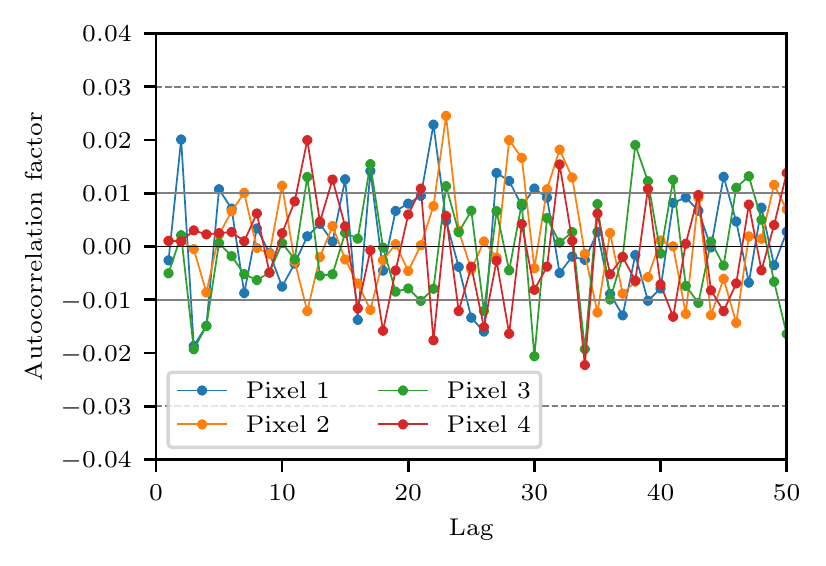}\label{sub:autocorrelation}}%
  \caption{(a) Probability distribution of the Pearson correlation factors measured between all pairs of pixels. The standard deviation $\sigma$ on the correlation factor is 0.01 which correspond to the uncertainty expected for the size of our data. (b) Autocorrelation of 4 pixels from the array. The solid and dashed grey lines represents respectively the confidence intervals of $\sigma$ and $3\sigma$.}
  \label{fig:correlations}
\end{figure}

In our model, we supposed that pixels are independent from each over (no crosstalk) and that the result of a pixel in one acquisition frame has no effect on the next frame. In order to validate these hypothesis, we acquired 10000 frames and we calculate the Pearson correlation coefficient $\rho_{ij}$ between pixels $i$ and $j$ and the autocorrelation coefficient $\rho_{i}(l)$ for pixel $i$ at lag $l$ :
\begin{equation}
\begin{aligned}
\rho_{ij} &= \frac{\left<\left(Z_t^{(i)}-\left<Z^{(i)}\right>\right)\left(Z_{t}^{(j)}-\left<Z^{(j)}\right>\right)\right>}{\sigma_i\sigma_j},\\
\rho_{i}(l) &= \frac{\left<\left(Z_t^{(i)}-\left<Z^{(i)}\right>\right)\left(Z_{t+l}^{(i)}-\left<Z^{(i)}\right>\right)\right>}{\sigma_i^2}
\end{aligned}
\end{equation}
where $Z^{(i)}_t$ is the value returned by the pixel $i$ at time $t$. Results are given in \cref{fig:correlations}. As we can see in \cref{sub:correlation}, the values of $\rho_{ij}$ are normally distributed around 0 and with a standard deviation of 0.01. This corresponds to the expected uncertainty of the measurements with a sample size of 10000. On \cref{sub:autocorrelation}, we plotted the values of $\rho_{i}(l)$ for 4 pixels on the CMOS array. For $l=1$, the autocorrelation coefficient is already in the uncertainty region due to our samples size and then fluctuates around 0 at all lags. These results validate the assumption made in our model that correlations are negligible and will not affect the entropy of the device.

\floatsetup[figure]{style=plain,subcapbesideposition=top}
	
	

\section{Quantum entropy estimation}

\begin{figure}[htbp]
	\includegraphics[width=\columnwidth]{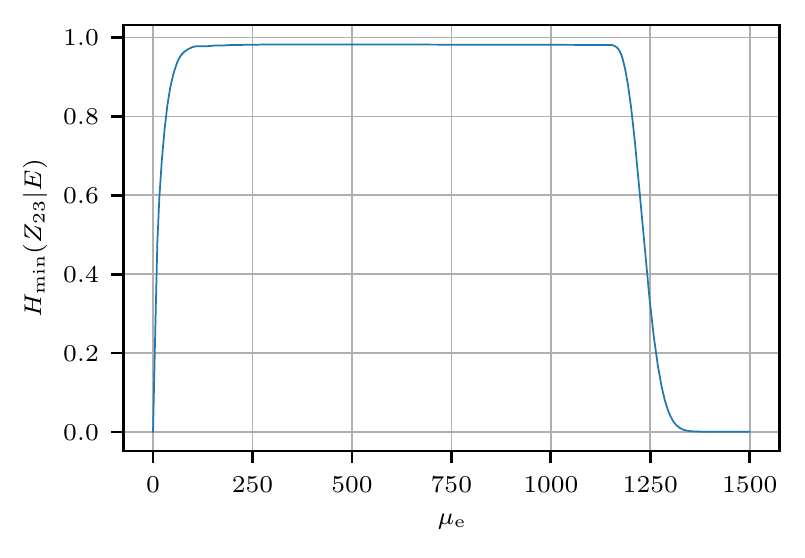}
	\caption{Quantum entropy as a function of the mean photon number simulated based on the classical noise characterization of for pixel 1.}
	\label{fig:final_entropy}
\end{figure}

Following the characterization of the chip, we can calculate the final quantum entropy of our two bits per pixel as a function of $\mu_\mathrm{e}$. Results are shown in \cref{fig:final_entropy}. As we can see, the quantum min-entropy is very close to its maximum value for a large range of $\mu_\mathrm{e}$ making it robust against fluctuations of the light intensity. It is also robust against small variations of the classical noise parameters whose effects only appear on the sharp edges of the curve. For $\mu_\mathrm{e}\in[500,750]$ which is the range where the chip normally operates, $H_{\min}(Z_{23}|E)$ is over 0.98 per bit which is a significant improvement compared to the 0.57 on average per bit measured in Ref.~\cite{Sanguinetti2014} for a specific intensity of the LED. However, with this device, we do not have access to the mean photon number arriving on each pixel to ensure we are in the optimal region i.e. 
\begin{equation}
\overline{H}_{\min}(Z_{23}|E) \geq H_{\min}^l
\end{equation}
where $\overline{H}_{\min}(Z_{23}|E)$ is the average min-entropy per pixel over the array and $H_{\min}^l$ is a lower bound on the entropy per pixel. If no control is implemented, fluctuations of the LED intensity or of the pixel efficiencies could lead to a degradation of the entropy. To make sure the chip is always providing the optimal entropy, 
we can define two thresholds on the ADC output $T^-$ and $T^+$ to record on each frame how many pixel outputs $n^-$ and $n^+$ were out of the interval $[T^-;T^+]$. If $n^\pm$ exceeds a predefined value $N^\pm$, it is registered as a failure and the frame is discarded. 

As we know the distribution of $Z$ for all pixels as a function of $\mu_\mathrm{e}$, we can therefore calculate the probability of failure $p_f = 1-\epsilon$ and the average min-entropy $\overline{H}_{\min}(Z_{23}|E)$ per pixel of one frame for any distribution of the light intensity over the array. For predefined values of $\epsilon$ and $H_{\min}^l$, appropriate parameters $T^\pm$ and $N^\pm$ can be found such that : 
\begin{equation}
	\mathrm{Prob}\left(\overline{H}_{\min}(Z_{23}|E)\leq H_{\min}^l\right) \leq \epsilon
\end{equation}
%

\floatsetup[figure]{style=plain,subcapbesideposition=top}
\begin{figure}[htbp]
  \sidesubfloat[]{\includegraphics[width=0.9\columnwidth]{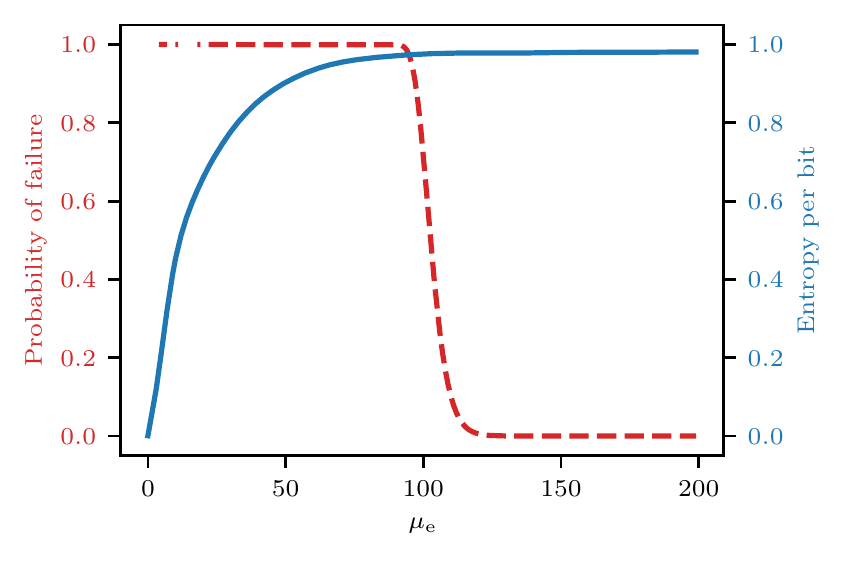}\label{sub:minus}}\\%
  \sidesubfloat[]{\includegraphics[width=0.9\columnwidth]{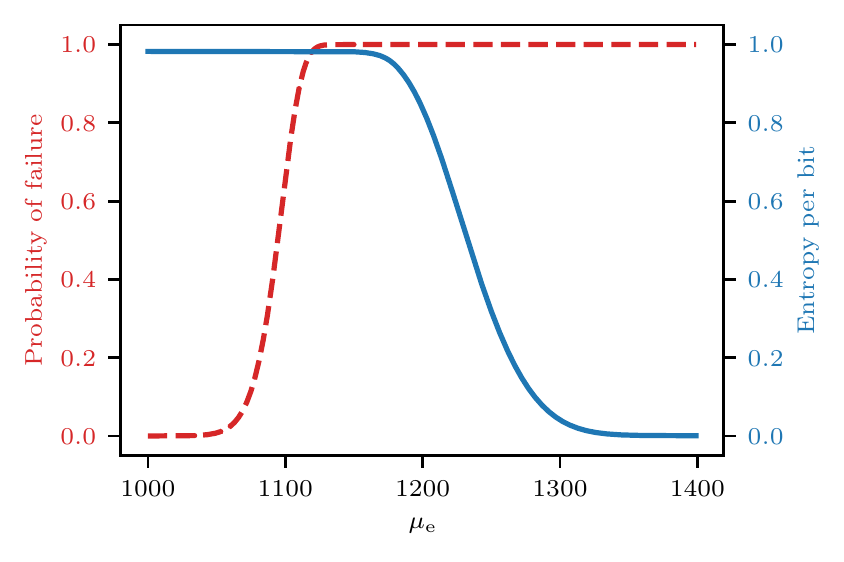}\label{sub:plus}}%
  \caption{Probability of failure and quantum entropy per bit of an array of 64 pixels uniformly illuminated as a function of the mean photo-electron number.}
  \label{fig:p_fail}
\end{figure}

As an example, we consider a chip with 64 pixels uniformly illuminated. In \cref{fig:p_fail} is plotted the probability of failure and the entropy per bit if the power of LED is drifting. The simulations were ran with $N^\pm=1$, $T^-=64$ and $T^+=940$. With this configuration, we can see that the entropy per bit is only dropping in the region were the failure probability is equal to 1. Other scenarios (e.g. one or several pixels losing efficiency) give similar results. This provides a strong indication that the chip can provide long-term robustness against LED failures ``in the field" because it will raise an alarm before the quantum entropy is even impacted.

\section{NIST tests}
	
The quality of our entropy source is assessed using the tests suite provided by NIST (details on the procedure can be found in \cite{NIST_SP800_90B}. The IID (independent and identically distributed) track of the test suite gives an entropy estimation of over 0.998 per bit for 10~Mbytes samples, using MCV (most common value) estimator. This value is higher than the 0.98 per bit given in \cref{fig:final_entropy} because the entropy test takes into account all sources of noise (quantum and classical) without distinction. If we run our simulations without considering that the classical is accessible to Eve, we obtain a value for the min-entropy of 0.999 per bit which is very close to the NIST result. This highlight an important advantage of our model compared to NIST entropy test. We can isolate the quantum contribution from the rest in order to calculate the quantum min-entropy.

\begin{figure}[!htbp]
	\includegraphics{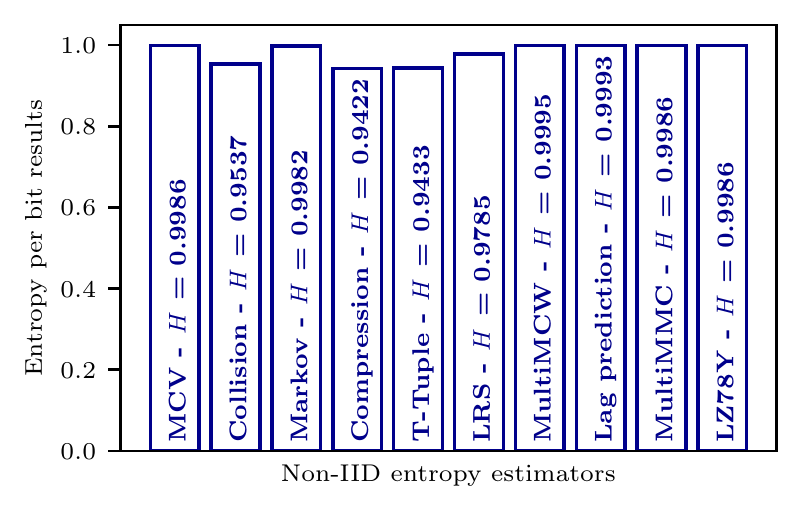}
	\caption{Typical results for the different entropy estimators on the NIST non-IID tests. The tests are carried out on 10~Mbytes samples.}
	\label{fig:nist}
\end{figure}

We also run the non-IID tests which consists of 10 different entropy estimators. Results are presented in \cref{fig:nist}. This approach is more conservative as it takes the lowest value of all the estimators and does not assume that the IID hypothesis is true. Nevertheless, this method gives for our chip an entropy value over 0.94 per bit. We can note that this value is lower than the one given by our model. This difference comes from how the tests are done. The entropy estimation is based on some statistical properties of a sample with finite size output by the device. Due to statistical fluctuations, the entropy estimated will be slightly different than its true value. We ran these tests with other entropy sources and with DRBG and the entropy value we obtained was always around 0.94 which tends to show that it is a limitation of the tests and not of the chip.

\section{Conclusion}
In this paper, we presented a physical model for the quantum entropy of the architecture on which the quantum random number generator of ID Quantique are based. With our model and after characterization of the device, we estimated that our chip can provide a quantum entropy of 0.98 per bit with a simple and low power consuming filtering of the bits. Finally, we show that the performances of the chip are robust against fluctuations over time making it suitable for mobile applications.


\acknowledgments
This project was funded from the European Union's Horizon 2020 programme (Marie Sk\l{}odowska-Curie grant 675662) and from the European Union’s Horizon 2020 research and innovation programme under grant agreement N$^\mathrm{o}$ 820405. We thank Florian Fr\"owis and Hyoungill Kim for helpful discussions.


\bibliography{library}

\end{document}